# Advances in dynamic AFM: from nanoscale energy dissipation to material properties in the nanoscale

Sergio Santos[1], Karim Gadelrab[2], Chia-Yun Lai[1], Tuza Olukan[1], Josep Font[3], Victor Barcons[3], Albert Verdaguer[4], Matteo Chiesa[1,5]

[1]Arctic Renewable Energy Center (ARC), Department of Physics and Technology, UiT, Norway

[2]Department of Materials Science and Engineering, Massachusetts Institute of Technology, Cambridge, MA 02139, USA[2]

[3]Departament d'Enginyeria Minera, Industrial i TIC, UPC BarcelonaTech, 08242 Manresa, Spain

[4]Institut de Ciència de Materials de Barcelona (ICMAB)
Campus de la UAB, 08193 Bellaterra, Spain

[5]Laboratory for Energy and NanoScience (LENS), Khalifa University of Science and Technology, Masdar Institute Campus, Abu Dhabi, UAE

Abstract

Since the inception of the atomic force microscope AFM, dynamic methods have been very fruitful by establishing methods to quantify dissipative and conservative forces in the nanoscale and by providing a means to apply gentle forces to the samples with high resolution. Here we review developments that cover over a decade of our work on energy dissipation, phase contrast and the extraction of relevant material properties from observables. We describe the attempts to recover material properties via one dimensional amplitude and phase curves from force models and explore the evolution of these methods in terms of force reconstruction, fits of experimental measurements, and the more recent advances in multifrequency AFM.

# I. Introduction

The applications of dynamic Atomic Force Microscopy dAFM have been expanding from its inception from surface science and the nanometric characterization of surfaces to the fields of biology[1], nanofabrication and nanomanipulation[2]. From the point of view of the enthusiastic researcher, dynamic methods broaden the field of AFM in that they present the possibility to engage in rich nanomechanical, chemical and biological investigations in a myriad of ways that involve interdisciplinary research from calculus[3-8] and combinations of electronics, control, dynamics, and simulation [9-14] to biology[15-17] and medicine[2, 18, 19]. Not only the dynamics of the micromechanical cantilever offer a platform for research but the tip-sample interaction provides an opportunity to investigate the non-linear interactions of a micro-mechanical system[20-22]. Furthermore, both the tip's and the sample's response occur in a nanometric "volume" of interaction where the concept of force and Newtonian physics can be pushed to the limit[23]. In 1998 an analytic expression that relates phase contrast, i.e. a channel characteristic of dynamic methods, to energy dissipation was presented by two groups independently[24, 25]. Cleveland et al. emphasized in their work that the very name of the field puts an emphasis on force over energy. However, adopting the energy perspective provided the means to map intrinsic materials properties with minimal need for data postprocessing. Both expressions are equivalent and contain information about the interaction occurring at the frequency the cantilever is driven at. In standard monomodal dAFM energy enters the cantilever at a frequency ω (drive frequency) only and energy dissipates to the tip-sample junction, the cantilever and the medium[25]. In approximately the same period, the conservative force was also investigated[5] by exploiting what was considered to be a negligible anharmonicity of the tip motion[26]. This simplification led to relatively robust analytic techniques to monitor the interplay of energy transfer[25] and forces[5] between the drive and the microcantilever-sample system. The cantilever response in the frequency domain was also exploited in the AFM field but it was arguably starting with the work of Stark et al.[27] that the harmonic spectrum was considered for imaging, spectroscopy and extracting general information about the sample in dAFM[12]. The addition of a second drive frequency near the second flexural mode introduced extra contrast channels [14, 28], the possibility to expand the mathematical relationships between force and material properties[3, 4, 13], improve contrast[29] and enhance the coupling of forces[4, 30, 31] and energy transfer [32-34] between harmonics and eigenmodes. Here we provide a review of the developments that our group and collaborators have brought in the field of nanoscale energy dissipation over the past decade and discuss it in the context of the development of dynamic AFM. Since we focus this review mostly on the work we have carried out over the years, most of the discussion focuses on ambient dAFM.

## II. Area of interaction

We start by discussing the area of interaction between the tip's and the sample's surface. If we are speaking about an oscillating system (dAFM) possibly interacting non-linearly via a surface force, we must understand that one thing is instantaneous power, or instantaneous energy dissipated, and another thing is the average energy or power dissipated per cycle. It is customary to write average quantities with either overbars $\overline{E_{dis}}$ [24], as expectation (average) values[23], i.e. <$E_{dis}$>, or otherwise state that we are dealing with averaged quantities over many cycles, i.e. the steady state[25]. An analysis of the area of interaction is conceptually equivalent since forces, let those be conservative or dissipative, are effective over an area and not at a single point. In short, energy might be unevenly distributed over the area[23]. This concept is seldom discussed in the literature and we want to clarify that the more localized the force, let that be conservative or dissipative, the higher the limit of resolution even when exploiting a given feedback system. We define[23] a mean areal density value <ρ> for the energy of interactions <E> over an effective area <S> as

$$<\rho> = \frac{<E>}{<S>} \tag{1}$$

The above expression is analogous to thinking in terms of average mechanical stress over a surface rather than stress at a point. It further helps us relate the concept of area on interaction S to energy dissipation $E_{dis}$ and a discussion on how to extract material properties from surfaces.

Figure 1 has been reproduced from an early (1993) work in AFM[35] where the effects of the tip on contrast, i.e. artifacts, were investigated. It is arguably easy to trivialize this issue, and therefore also the issue of resolution or the acquisition of material properties from a surface's spot[36], and reduce the interpretation to the fact that an effectively spherical nanometric, i.e. finite, tip of radius R interacts with a surface. It follows from this argument that R together with the roughness of the surface controls resolution and the capacity to quantify the properties of a material (Figure 1). On other hand, here we want to discuss this problem more thoroughly and show that resolution is governed by A) geometry in a way that is subtly coupled with material properties and the character of the forces, and B) the heterogeneity of the surface in terms of forces and properties.

## II.A The problem of geometry

First, the tip-sample interaction is a geometrical problem[23, 35, 37, 38]. That is, the geometry - including the distance from the tip to the surface – controls the force (Figures 1 and 2).

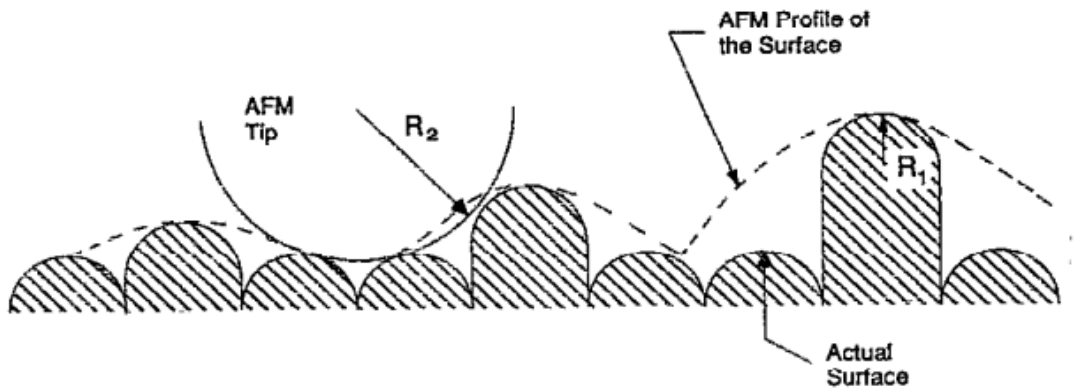

Figure 1. Sketch of a tip of effective radius $R_2$ scanning over a rough surface with features of radius $R_1$ and a given height.

The fact that tip-sample force expressions account for the geometry of the tip and the sample via extensive parameters makes this clear (see expression below and in the literature throughout). For example, in the short range, i.e. where mechanical contact occurs, contact mechanics models such as the Deraguin-Muller-Toporov (DMT)[39] or similar[40] are invoked to exploit the relationship between the contact radii $r_c$ and tip radius R [23],

$$r_c = [R\delta]^{1/2} \qquad d<a_0 \qquad (2)$$

where r stands for contact radius, the subscript c stands for contact, the tip-sample distance d at d=$a_0$ is an intermolecular distance that indicates the point of mechanical contact between the sample and the surface and δ is the indentation[41]. Computing the contact area here is trivial, i.e. πRδ. In the "long range", i.e. d≥$a_0$, the concept of "contact area" is particularly ambiguous. An expression such as (1) would be convenient but long-range forces might interact over large structures. In 2011 we developed the concept of effective radius and area of interaction for van der Waals (vdW) type of forces[23, 42, 43]. We termed the area S to distinguish it from the Amplitude of oscillation A. To this end we exploited what is typically known in AFM as the Hamaker expression (Eq. 2). The Hamaker approach is typically exploited to model long-range vdW interactions[41] where a triple volume integral leads to the relatively compact expression for the force F [44]

$$F_{nc} = \frac{RH}{6d^2} \qquad d \geq a_0 \qquad (3)$$

where the subscript nc stands for non-contact and H is the Hamaker coefficient that contains information about material properties and electronic structure[40, 44]. We note from Figure 2a, adapted from Ref. 23, that all the atoms in the sphere (tip) interact with all the atoms in the infinite plane or surface[23, 42]. Nevertheless, the actual interaction can be reduced to an arbitrary fraction p of the total energy. This allows us[23] to write a given effective radius r(p) and area S(p) correspondingly. This is indicative that we are accounting for a fraction p of the interaction only (Figure 2). We further showed that the area <S> can be conveniently expressed in terms of R and the tip sample distance d as

$$r_{nc}(p) = aR^{1/3}d^{3/5} + bR \qquad d \geq a_0 \qquad (4)$$

where a and b need to be given the appropriate units to satisfy the units of $r_{nc}$ – see Ref. 23 for actual values. It is possible that these expressions can be further simplified. To our knowledge however, there is no advance in the literature in this respect.

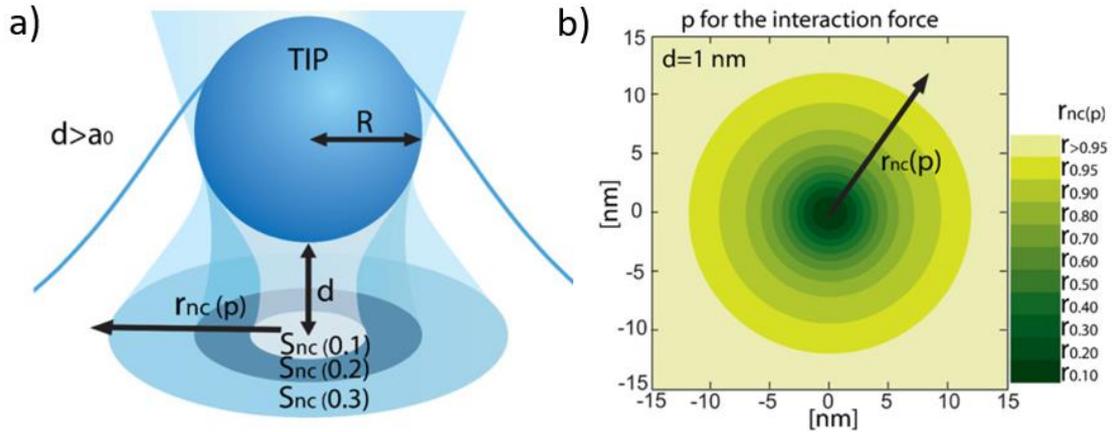

Figure 2. a) Tip of an AFM modelled as a sphere of radius R at a distance d from a surface. b) Effective radius of interaction in the non-contact region, $r_{nc}$, at a distance d=1 nm. Most of the interaction is due to the atoms right below the tip but the interaction with atoms nm away is not negligible.

The fact that most of the interaction concentrates in a nanometric region, i.e. $r_{nc}$~ nm, at distances where the AFM typically operates, i.e. d≈ 1nm, is consistent with the possibility of achieving high lateral resolution with the AFM even with these "dispersed" forces, i.e. London Dispersion Forces (Figure 2b). The concept described with the help of Figure 2 already shows that the tip does not need to "mechanically touch" the surface to sense it or be affected by it. Therefore, the concept of geometry in AFM goes far beyond understanding mechanical gaps or overlapping (Figure 1). Expanding the concept of area should lead to stronger mathematical formulations of the limits of resolution[23, 45, 46]. For example, we used[46] the concept of spatial horizon (SH) to establish the distance at which the tip "sees", i.e. the feedback system detects the presence of, an atom (Figure 3 adapted from Ref. 46). In the same work we compared the SH in terms of amplitude and frequency feedbacks, these still being the most common forms of feedback.

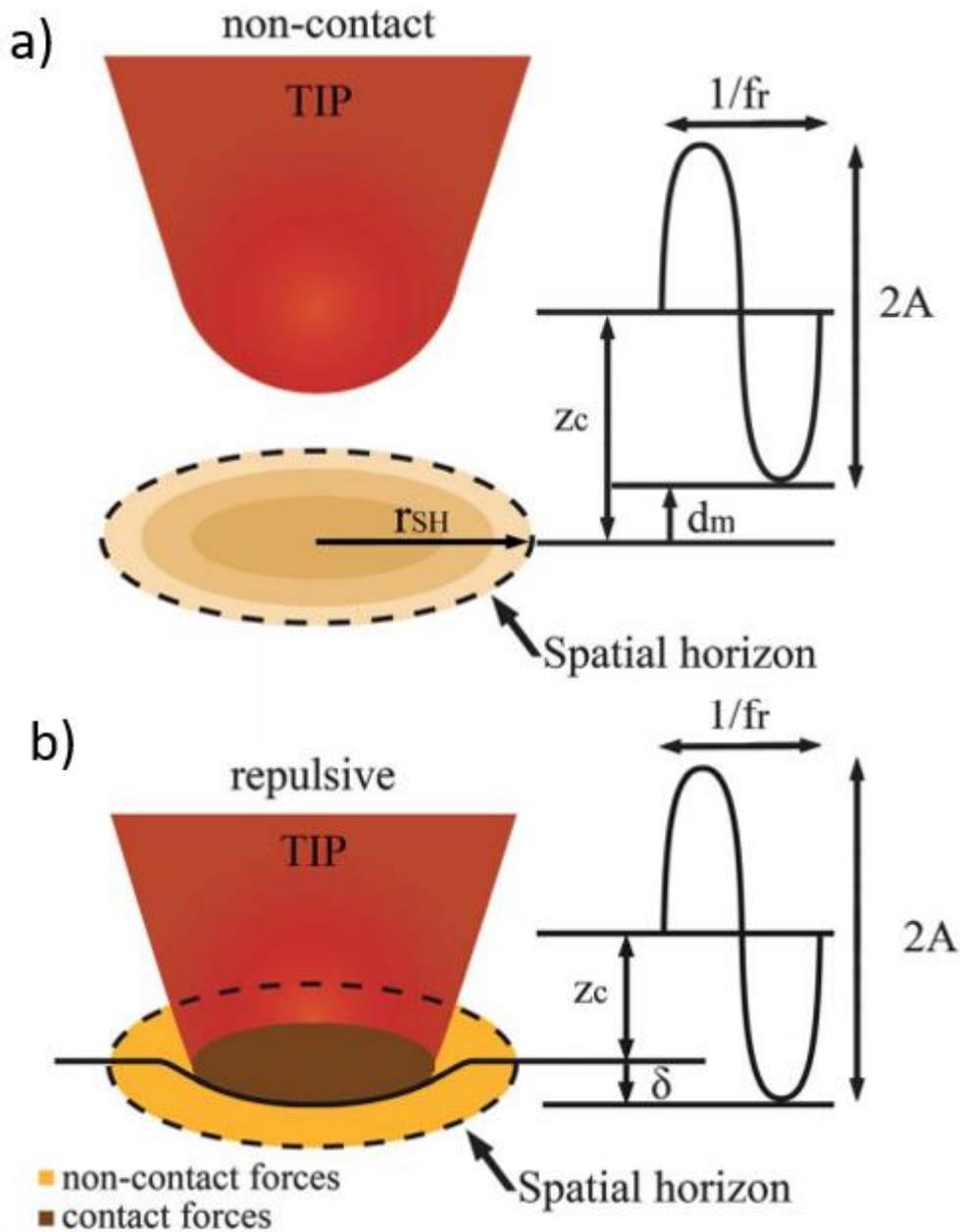

Figure 3. a) Scheme of a tip vibrating in the non-contact mode where $d_m > a_0$ and $\delta=0$. Here $d_m$ is the minimum distance of approach during one cycle, $a_0$ is the intermolecular distance that physically implies that matter interpenetration cannot occur and $\delta$ is the sample indentation. The spatial horizon (SH) is thus affected by long range forces only. The interactions occurring between the tip and the sample's atoms lying beyond the boundary established by the SH do not sufficiently affect the dynamics of the cantilever for the feedback to detect them. b) Scheme of a tip vibrating in the repulsive regime where intermittent mechanical contact occurs, $d_m < a_0$ and $\delta > 0$. The SH in this case is affected by both short range and long-range forces (adapted from Ref. 46).

The concept area of interaction S, or spatial horizon SA, has other non-obvious consequences when measuring the apparent height of nanostructures since the tip might "see" different structures when generating contrast[45, 47, 48]. For example, does the contrast belong to the surface or to the nanostructure on top of the surface being imaged? We showed that contrast arises from both contributions[45, 49]. For example, we discussed how a tip R=5nm positioned on top of a dsDNA molecule would be affected by contributions from the surface and several minor and major grooves forming de characteristic helices of dsDNA (Figure 4 adapted from Ref. 49).

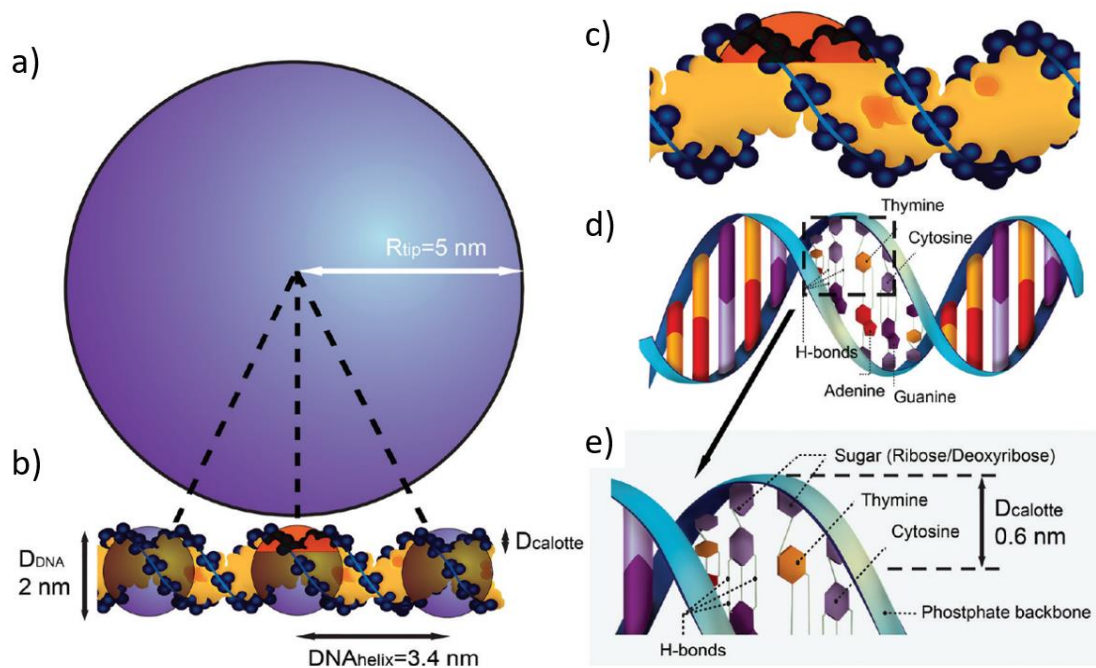

Figure 4. a) Scheme of the geometry of the tip−sample interaction drawn to scale. The tip is positioned on top of the highest parts of one of the helices, i.e. the minor groove, with a helical pitch of 3.4 nm. Each groove, including the surface, accounts for part of the net tip-sample force. b) Schematic diagram of a dsDNA molecule and (c and d) chemical composition and distribution. Adapted from Ref. 49.

In Figure 5 we reproduce the schematics (a-d) and experimental outcomes of imaging a dsDNA molecule on a mica surface in dAFM (adapted from Ref. 45). Figure 4c illustrates what is expected, i.e. 2 nm of apparent height and width, and Figure 4e together with Figure 4f show an actual image of the molecule and the measured height and width correspondingly. The molecule "loses" 50% of apparent height and displays a much larger apparent width. These are some of the consequences of dealing with an area of interaction or SH. To sum up this section, geometry alone might affect resolution, height and contrast on account of force-distance dependencies. We next discuss how material properties, and in particular heterogeneity in material properties, also lead to a coupling between different contributions to the area of interaction and therefore resolution and contrast.

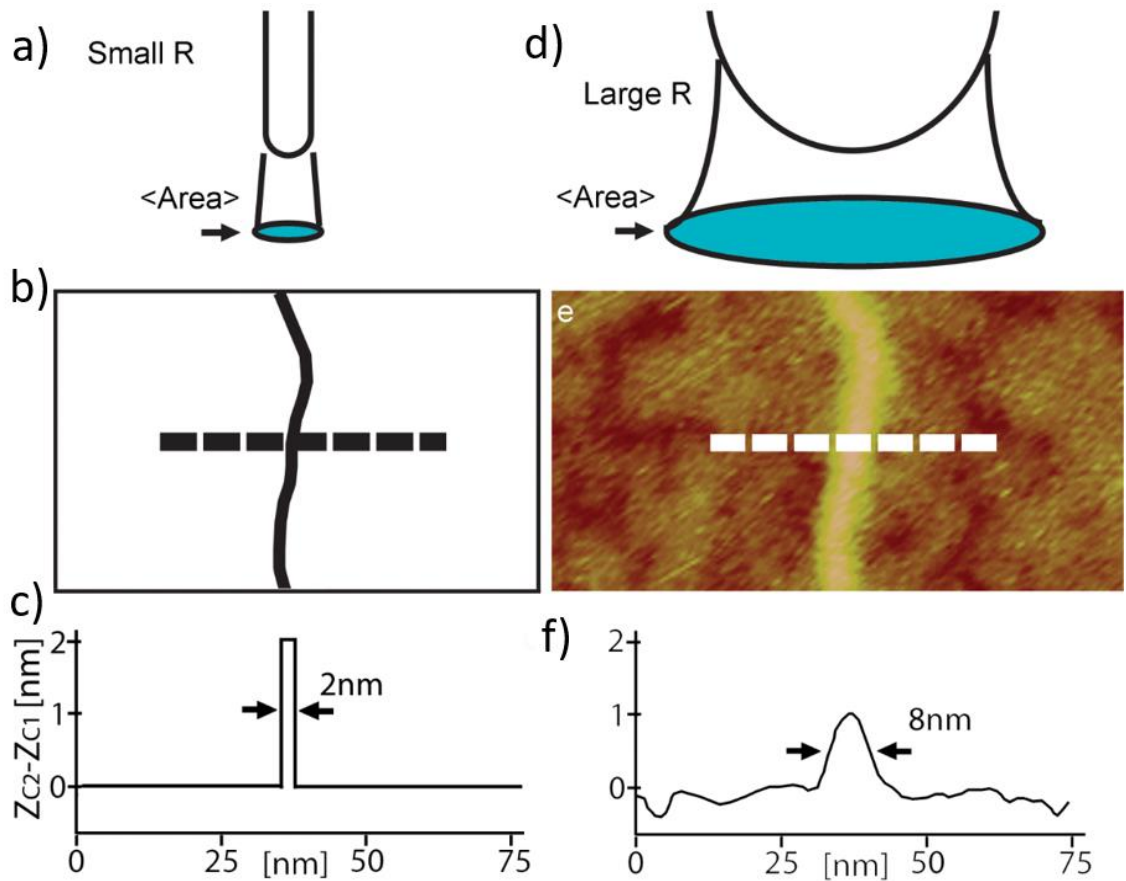

Figure 5. Scheme of a hypothetical point area versus experimental outcomes where the area is finite. Adapted from Ref. 45.

## II:B The problem of heterogeneity, contrast, and material properties

As a second point we indicate that the interaction is controlled by the properties of the interacting materials. This is also clear from the fact that force and energy expressions include the properties of the interacting materials. This implies that resolution, including the recovered height of the surface structures[45, 50], is also governed by such contrast since the resulting different forces might have 1) different distance dependencies or 2) different absolute values at a given distance. Such differences indirectly affect the effective area of interaction. In general, topography might be coupled with chemistry while heterogeneity in terms of material properties couples with topography in a non-trivial way. To these considerations we must add the problem of always dealing with a finite area of interaction where contributions to the force add (Figure 6 adapted from Ref. 78. ).

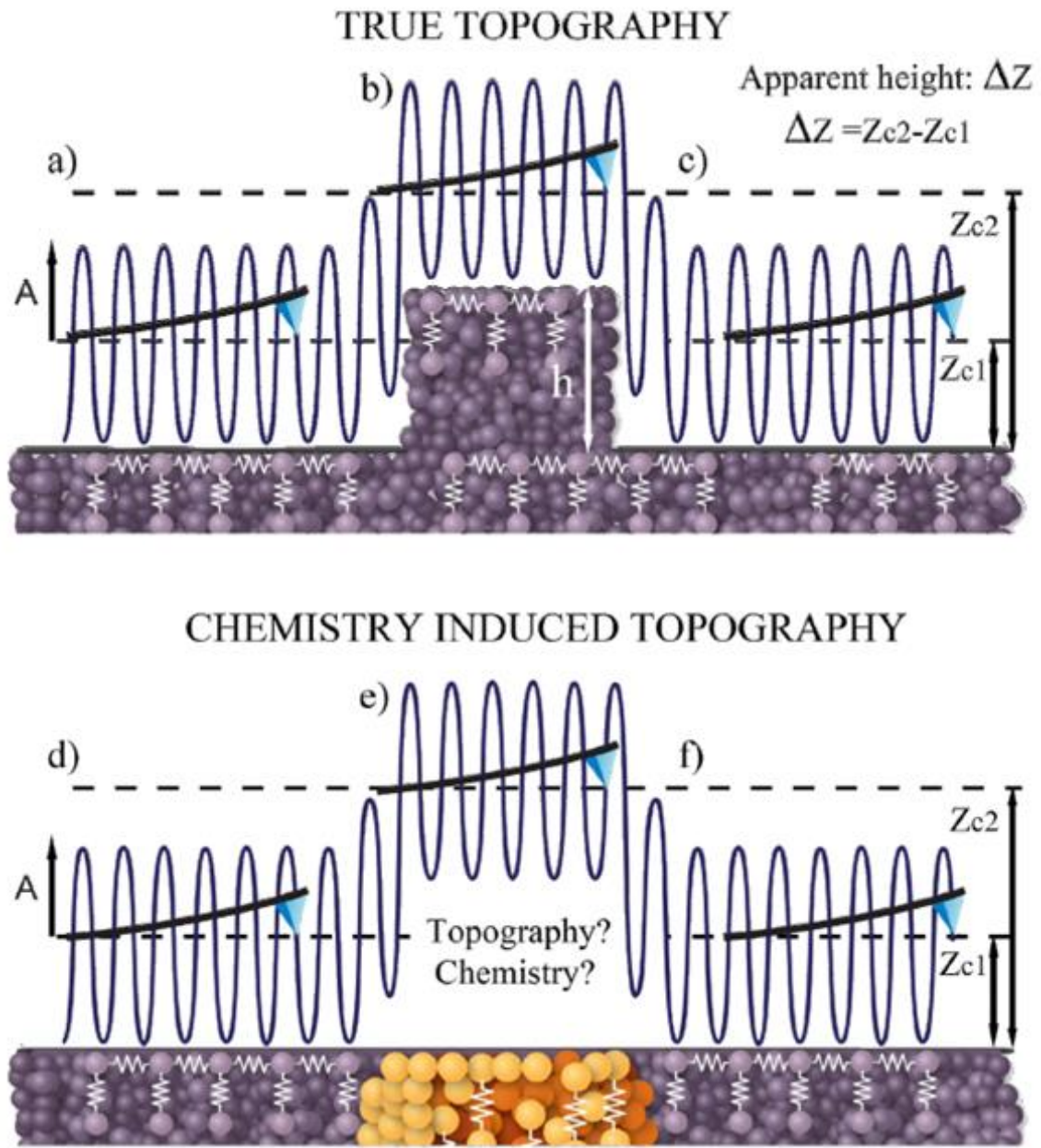

Figure 6. a–c) Illustration of a cantilever oscillating above a surface and recovering the true height h when there are is no compositional heterogeneity or chemical variations. d–f) Topographical variations induced by chemical or other compositional heterogeneity. Adapted from Ref. 78.

There is another important issue regarding material properties since these are typically considered to be intensive parameters. In this respect, another two points need clarification. On one hand, the nanoscale regime might affect the behaviour of the material[37, 51] and being a probe-based microscope, in AFM the tip must be considered with care (Figure 7 adapted from Ref. 51). For a given cantilever-sample set of parameters, the tip radius might lead to a quite different interaction as shown in the illustration of Figure 7. On the other, the dynamics of the cantilever, such as at the frequency of interaction[52], might also affect the quantifiables.

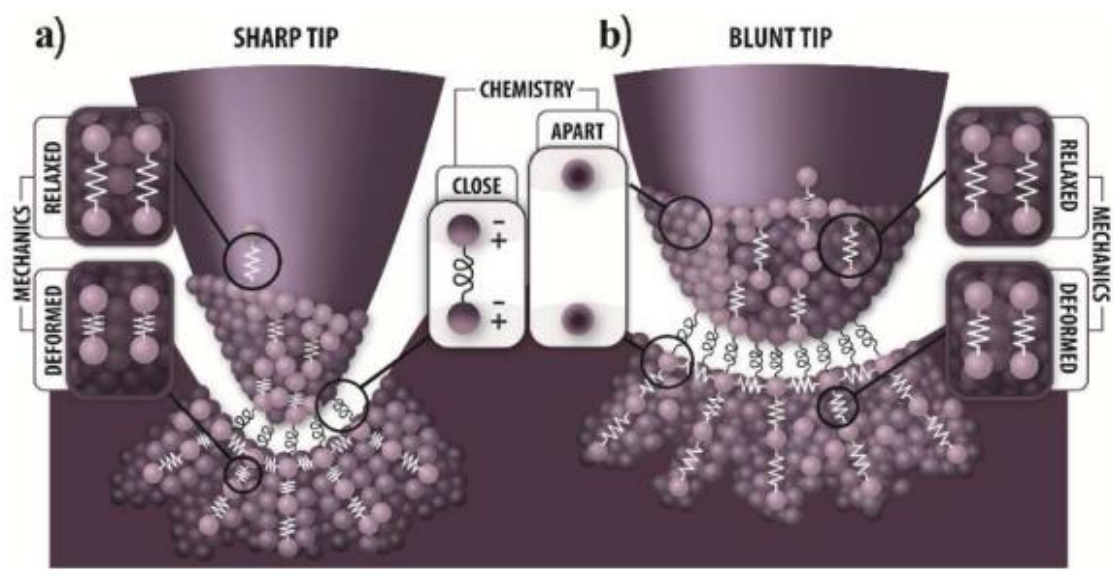

Figure 7. Scheme of the possible interactions occurring between an AFM tip and a surface when a) the tip is very sharp and (b) when it is blunter. Adapted from Ref. 51

## III. Phenomenal description of energy dissipation and mechanisms

To understand the dynamics of the cantilever and how dynamics are controlled by forces, and therefore provide a channel for characterizing and quantifying materials, the equation of motion is typically invoked. After some assumption of linearity and reducing the cantilever to its first mode[53] we can write

$$m\ddot{z} + \frac{m\omega_0}{Q}\dot{z} + kz = F_{ts} + F_o\cos\omega t \qquad (5)$$

where ω is the drive frequency, m is the effective mass, Q is the quality factor and k is the spring constant. These 4 parameters can be calibrated when imaging[54]. The drive force $F_0$ can be derived from standard vibration theory and gives $F_0=kA_0/Q$ where $A_0$ is the oscillation amplitude of the unperturbed cantilever.

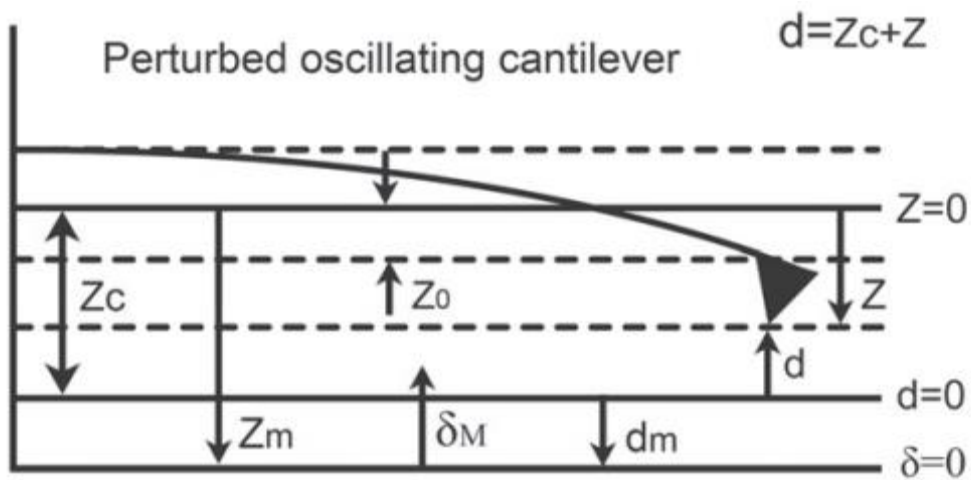

Figure 8. Schematic diagram of the relationship between the different geometrical parameters typically employed in dAFM. Adapted from Ref. 62.

The response in terms of the first harmonic only is z=Acos(ωt+ϕ) where A is the oscillation amplitude and ϕ is the phase shift between drive and response. The somewhat standard terminology of the geometry of the interaction is shown in Figure 8 (Adapted from Ref. 62). We note that the tip-sample distance d is related to the tip position z relative to the equilibrium point of the cantilever $z_c$, d=$z_c$+z. The tip-sample force $F_{ts}$ is clearly affecting the dynamics by acting as a "drive" albeit one through which energy can only be dissipated. We note however that even via elastic interactions, that is, even if energy is not lost in the tip-sample junction, the amplitude A can decay. A typical question here is: where does the energy go if not in the tip-sample interaction and inelastically? The answer is that the medium dissipates energy through Q. So even when energy is not dissipating in the tip-sample volume (Figure 9) in some way, the non-linearity of the interaction leads to energy transfer to higher harmonics and modes and the medium dissipates energy through those[27, 33, 53]. We note that considering the response z in terms of the full set of harmonics shows that frequencies other than that of the drive ω follow[53]. Since these frequencies have no drive, they must ultimately get their energy from the drive ω. This phenomenon is discussed later in the section dealing with energy transfer, but Figure 9 shows the concept "effective volume of interaction" ΔV or tip-sample junction. In this respect, in this section we discuss the energy dissipated from the dynamics of the cantilever "to" the tip-sample junction. Energy dissipated in this way is typically[24, 25] termed $E_{dis}$. More thoroughly force expressions related to each dissipative phenomenon do work against the tip-motion and at the drive frequency one gets that this energy can be directly computed from observables as

$$E_{dis}(\omega) = \frac{\pi k A_0 A}{Q} \left[ sin\phi - \frac{A}{A_0} \right] \quad (6)$$

It follows directly from (6) that "conservative" interactions, i.e. written in brackets because energy might still dissipate at other frequencies, are those that satisfy the condition

$$sin\phi - \frac{A}{A_0} = 0 \qquad (7)$$

It is remarkable that one needs to know nothing of the interaction, irrespective of how complex, in order to check if energy dissipates onto the sample in the tip-sample junction. We already noted[55] however that $E_{dis}$ and other averaged quantities might be insufficient to fully characterize the fast dissipative processes occurring in the nanoscale, i.e. time averages in AM AFM lie in the range of microseconds and are too long with respect to the speed of propagation of atomic dissipative processes in the nanoscale. Thus, we can proceed and consider dissipation either classically (Figure 9a) or discretely (Figure 9b) - adapted from Ref. 55. It is worth noting that in standard dAFM operation $E_{dis}$~ 1-100 eV[43]. Since we have atomic events in the interaction which require energies in the range of ~ eV, these energies might be sufficient to perturb single atoms and these perturbations might lead to a transfer of momentum, rotation, or, in general, a transfer of kinetic and potential energy of vibration from the drive to the junction. A scheme illustrating possible dipole-dipole bond formation and rupture is shown in Figure 9b.

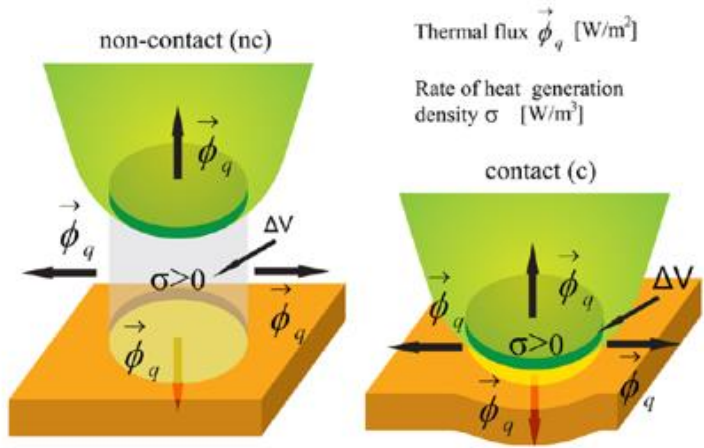

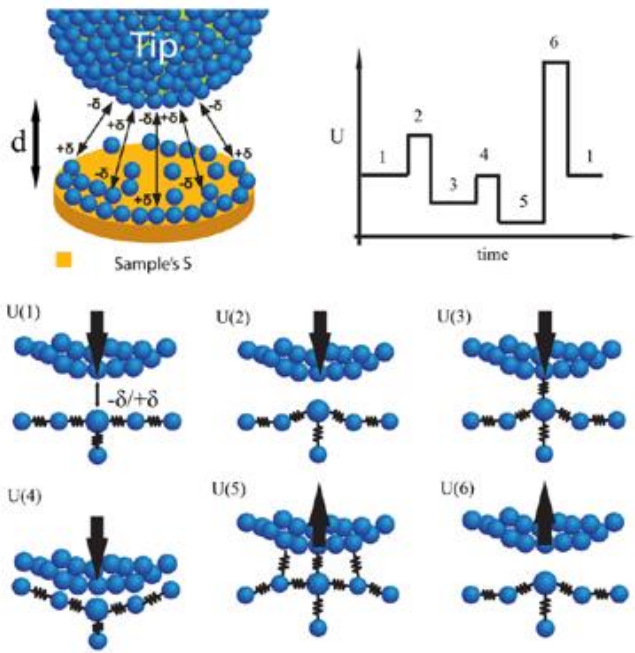

Figure 9. a) Schemes of the instantaneous volumes of interaction 1V where (a) non-contact and contact forces are present. If the areas and volumes of interaction are known, the rate of heat generation σ and the thermal flux can be obtained. B) Schematic of the tip–sample interaction occurring in an area S and inducing dipole–dipole forces amongst others. The energy diagram for the processes occurring in the interaction and where the energy of the interacting atoms is expressed as U. Each step is depicted in the schemes as either kinetic, potential or bond energy. Adapted from Ref. 55.

At the phenomenal level however and continuing with a classical approach that allows us to push the concept of force and Newtonian physics to the limit, we can divide dissipative mechanisms into 1) hysteretic and 2) viscous. The first relate to history-dependent interactions while the second belong to the domain of velocity dependent forces. López-Guerra and Solares[52] have recently described in considerable detail how in real situations a combination of hysteresis and viscosity, as a function of drive frequency, should be considered. Here we proceed to describe these two concepts with the help of the illustration in Figure 10 – adapted from Ref. 23. Viscosity in both the nc (Figure 10a) and the c (Figure 10b) regions oppose the motion of the tip and are not only position dependent but also velocity dependent. Informally this is the typical phenomenon of "friction" that increases with velocity. Since friction itself is not a fundamental force but depends on adhesion, atom inter-locking, contamination, surface roughness etc. Figure 9b provides a better picture of the possible mechanisms involved. More information was given in our work. Finally, while viscosity might depend on the direction, i.e. tip-approach or tip-retraction, history dependent dissipative mechanisms are typically termed hysteretic (Figure 10c).

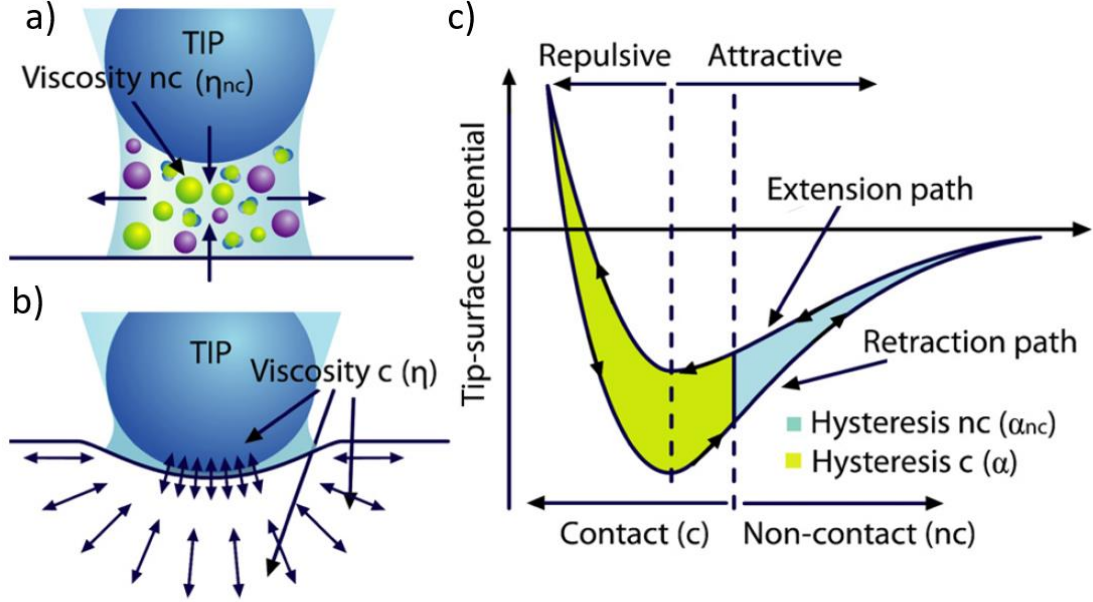

Figure 10. Illustrations depicting the mechanisms of viscosity in a) the nc region and b) the c region. c) Hysteretic processes are many times characterized by an increment in force on retraction, as the tip moves away from the surface, as compared to extension. Adapted from Ref. 23

## IV. Energy dissipation models and integral expressions

Considering the illustration in Figures 9 and 10 we now write down some simple but standard expressions for viscous and hysteretic forces typically used in the literature and added to the equation of motion in (5). In the short range viscosity has been modelled by assuming a Kelvin-Voigt material by some[56, 57], including us[58, 59], as follows

$$F_{ts} = -\eta_c r_c \dot{z} \qquad d<a_0 \qquad (7)$$

in this expression the "intensive"[52] parameter that contains information about the material is the viscosity of the sample $\eta_c$. We recall that the material that the tip is made of, together with its size, and the operational parameters, etc. might also play a role in $\eta_c$ but this is not considered in (7). In the long range we take the same concept and write

$$F_{ts} = -\eta_{nc} r_{nc} \dot{z} \qquad d \geq a_0 \qquad (8)$$

where $\eta_{nc}$ is the viscosity where no mechanical contact occurs, i.e. in the vdW range. We note that energy dissipated via (8) is not equivalent to energy dissipated to the medium via parameters such as the Q factor since here it is the presence of tip-sample forces that lead to dissipation and not the medium alone. Similarly long[60] and short range interfacial forces can be written as

$$F_{ts} = -\alpha_c F_{AD} \qquad d<a_0 \quad \text{and} \quad \dot{d}>0 \qquad (9)$$

$$F_{ts} = -\alpha_{nc} \frac{F_{AD} a_0^2}{d^2} \qquad d \geq a_0 \quad \text{and} \quad \dot{d}>0 \qquad (10)$$

where $F_{AD}$ is the force of adhesion or

$$F_{AD} = -\frac{RH}{6a_0^2} \qquad (11)$$

and $\alpha_c$ and $\alpha_{nc}$ are the coefficients of hysteresis that dictate how much larger is the force during tip-retraction than it is during tip-approach in each force regime. The fact that these forces act only when $\dot{d}>0$ indicates that they are hysteretic in nature, i.e. they do work even though they have a negative sign and do not depend on velocity. The integral form, i.e. the energy dissipated on average per cycle due to each contribution is

$$E_{ts} = \oint F_{ts}\, dz \tag{12}$$

The integrals for each expression in terms of the relevant parameters give[61, 62]

$$E_{ts}(\alpha_c) = \alpha_c F_{AD} \delta_M \tag{13}$$

where $\delta_M$ is the maximum indentation in the cycle provided there is indentation. This expression can also be written in terms of the difference in surface energy during approach and retraction[59].

$$E_{ts}(\alpha_{nc}) = \alpha_{nc} F_{AD} a_0^2 \frac{2A}{d_m(2A+d_m)} \tag{14}$$

where $d_m$ is the minimum distance of approach. If the oscillation amplitude A>>$d_m$ (a condition that is relatively easily satisfied), then

$$E_{ts}(\alpha_{nc}) \approx \alpha_{nc} \frac{F_{AD} a_0^2}{d_m} \qquad \text{A>>}d_m \tag{15}$$

Finally, for the short-range viscosity

$$E_{ts}(\eta_c) = \eta_c \frac{\sqrt{2}}{4} \pi \omega (RA)^{1/2} \delta_M^2 \tag{16}$$

Where all parameters have been determined. We leave the long-range viscosity term unsolved but note that the expression that gives the radius in Eq. 4 as a function of tip radius R is solvable for the viscosity in the non-contact region. We leave other dissipative mechanisms in ambient conditions such as the capillary interaction unexplained but refer the reader to the literature[63-65].

## V. Disentangling mechanisms and properties from observables

There are several advantages in writing simple expressions such as those in (7) to (10). In particular numerical simulations can be exploited to disentangle the contributions. Garcia et al. developed a method in 2006 to disentangle surface energy hysteresis, long-range interfacial interactions and viscoelasticity mechanisms exploiting amplitude versus distance curves and the expression of energy dissipation in (6).

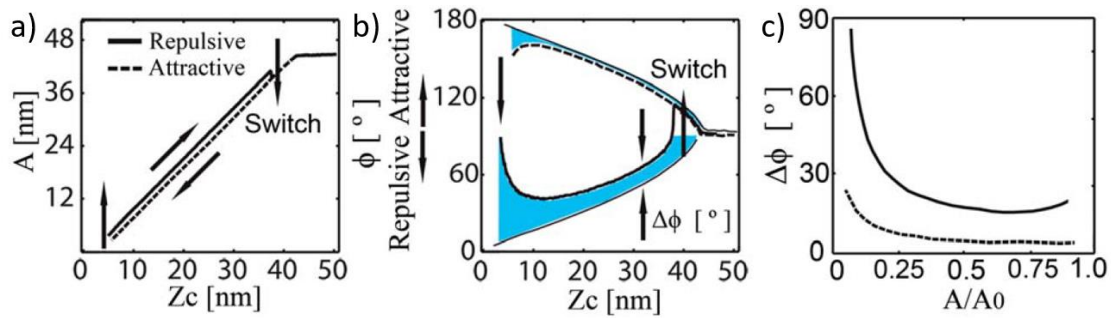

Figure 11. Experimental (a) amplitude and (b) phase distance curves obtained on a mica sample at 50–70% RH. The conservative phase branches are plotted with the thin continuous lines whereas the experimental phase branches where dissipation occurs are plotted with thick dashed (attractive) and thick continuous (repulsive) lines. (c) We term the difference between these two branches as the phase difference $\Delta\phi$. Adapted from Ref. 61.

Examples of experimental amplitude A and phase $\phi$ distance curves are shown in Figure 11 (Adapted from Ref. 61). We note that during approach and retraction two branches are found. This is not an artifact but responds to the phenomenon of bi-stability[66-68], i.e. two solutions, the attractive and the repulsive branch, might follow when vibrating near a surface. The difference in phase $\Delta\phi$ between what would be expected if there were no inelastic interactions Eq. (6) and the experimental phase are shown in Figure 11c and Figure 12 (Adapted from Ref. 58). The authors calculated the ratio of the difference, i.e. derivatives, in $E_{dis}$ with amplitude $A/A_0$. In 2012 we proposed [58] an alternative method that included viscosity in the long range and did not involve derivatives since derivatives typically lead to higher levels of noise. The attractive and repulsive branches for the phase are shown in the illustration of Figure 12 for the a) non-contact and b) contact viscous and hysteretic interactions. The presence of a singularity in $\Delta\phi$ allowed us to define two shapes in amplitude versus distance curves at once to distinguish between viscosity and hysteresis (see Ref. 58). In the short range this signal was not sufficient but we realized multiplying it by $E_{dis}$ would enhance it without adding much noise, i.e. we defined[58] the $E_{dis}\Delta\phi$ signal or channel. In the same work we showed (Figure 13) how for a CNT on a quartz surface that in the attractive regime (Figures 13 a and b), i.e. where long range forces operate, the mechanism was hysteretic in agreement with the hypothesis of Garcia et al. that hysteresis was a dominating mechanism in this regime. In the repulsive regime (Figures 13 c and d), i.e. where short range forces operate, we found both hysteretic and viscous mechanisms on the CNT and hysteretic only on the quartz surface. In the CNT viscosity was prominent for large oscillation amplitudes in agreement with Eq. 16 where $A^{1/2}$ is a factor.

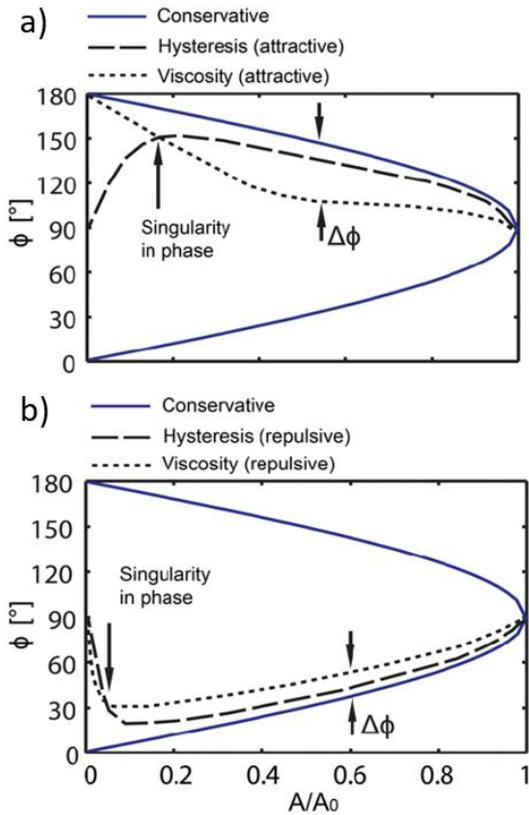

Figure 12. a) Scheme displaying the behaviour of the phase $\phi$ for the conservative interactions (continuous blue lines) and when dissipation occurs in the attractive regime through non-contact (nc) hysteresis (dashed lines) and viscosity (dotted lines); $\Delta\phi$ is defined as the difference in phase between the conservative and the dissipative branches. b) Scheme displaying the behaviour of $\varphi$ as in figure 1 but here dissipation occurs in the repulsive regime through contact hysteresis (dashed lines) and viscosity (dotted lines). Note that for these contact processes the concept of $\Delta\phi$ cannot be used to distinguish between them since both involve maxima when $A/A_0 \rightarrow 0$; note the behaviour in (a). Thus, the parameter $\langle E_{dis}\Delta\phi \rangle$ is used instead. Adapted from Ref. 58

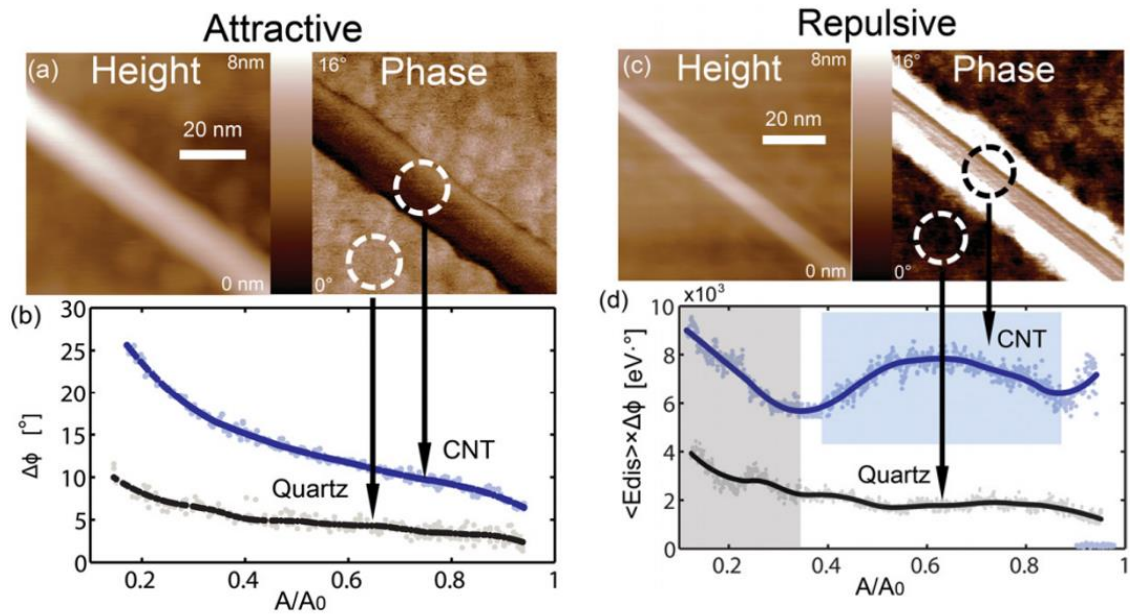

Figure 13. *a*) Height and phase contrast for a CNT on a quartz surface obtained in the attractive regime. *b*) Here $\Delta\phi$ displays no distinctive maxima at intermediate $\bar{A} = A/A_0$ values. Instead $\Delta\phi$ increases monotonically with decreasing $A/A_0$ throughout. This indicates that nc hysteresis dominates on both the CNT and the quartz surface. *c*) Height and phase contrast for the same CNT and quartz system obtained in the repulsive regime. (*d*) Here $E_{dis}\Delta\phi$ shows distinctive maxima at intermediate $A/A_0$ values only on the CNT (light blue-coloured region). Adapted from Ref. 58

In another work[59] we compared the method by Garcia et al. with our own experimentally and acknowledged that the mechanisms might be present but material properties could not be recovered through these methods. For this reason, we proposed a method to find a best fit for the dissipative parameters $\eta_c$ and $\alpha_c$. In principle the same could be done for the long-range parameters but we focused on relatively large amplitudes and the repulsive terms as Garcia et al. did in their work. We solved the problem of having insufficient equations by acknowledging the many points that are obtained in the amplitude and phase versus distance curves ~10-1000. Some results are reproduced in Figure 14 (adapted from Ref. 62) for a silicon tip interacting with a silicon sample. In Figure 14b we show a comparison between the method of Garcia et al. and ours. We found that the main dissipative mechanism was short range hysteresis as expected (Figure 14d). In the same work we disentangled the hysteretic and viscous components for the silicon, graphite, ferrite, and polypropylene samples interaction with a silicon tip (not shown). This work made us realize that even for a given tip-sample material, i.e. silicon-silicon, the controlling mechanism could vary from experiment to experiment. We partially solved this problem by monitoring on changes in tip radius R (see Figure 15).

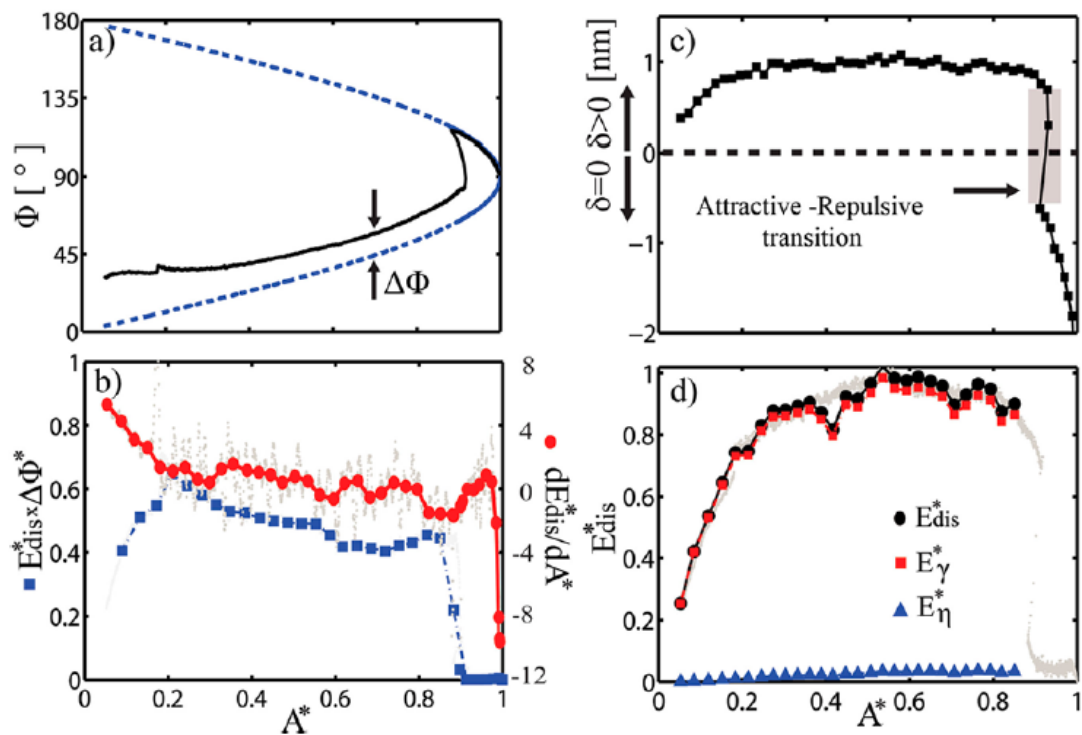

Figure 14. Experimental results for a silicon tip interacting with a silicon sample with a tip radius of R = 12 ± 2 nm as a function of normalized amplitude $\bar{A}$ a) Conservative

(dashed blue) and dissipative (continuous black) phase branches for the system where $\Delta\phi$ is the phase difference. b) Qualitative $d\overline{Edis}/d\bar{A}$ (red circles) and $E_{dis}\Delta\phi$ (blue squares) methods have been used to hint at the nature of the dissipative processes. In this case, surface energy hysteresis controls the interaction according to both methods. c) Reconstructed distance of minimum approach $\delta_M$. Where $\delta_M \geq 0$, tip−sample deformation occurs. Note that $\delta_M$ becomes positive only in the repulsive regime, i.e., after the force transition, implying that, in the attractive regime, there is no sample deformation. d) Decoupling of surface energy hysteresis $E_\gamma^*$ (red squares) from viscoelasticity $E_\eta^*$ (blue triangles) from total energy $E_{dis}^*$ (black circles). Adapted from Ref. 62

## VI. Force profile and energy dissipation

During the last years of the first decade of the century methods were developed to recover the conservative part of the force from amplitude versus distance curves in dAFM. Albeit in frequency modulation [5, 7] such methods were already more advanced back then than those in amplitude modulation[69-72]. Implementing these methods however sometimes requires simultaneous good calibration and experimental techniques[51, 73] that have been advancing over the years. In 2013 we first proposed[51] a method similar to those exploited to disentangle dissipative mechanisms from amplitude and phase curves that exploited force profiles instead (Figure 15, adapted from Ref. 51.). We note that the integral expressions from (13) to (16) depend on knowing either $F_{AD}$, $\delta_M$ or both. These values can be obtained directly from the force profiles as shown in the figures. We compared simulations (left panel in Figure 15) with experiments (right panel) and showed that the $E_{dis}\Delta\phi$ channel behaved very distinctively for viscous and hysteretic short range mechanisms. Remarkably, the same system would display the behaviour of one or the other mechanism depending on the sharpness of the tip. This is consistent with the illustration in Figure 7 and our interpretation there, i.e. everything else unchanged, a blunt tip will cover a larger area of interaction implying that hysteretic dissipation will increase. Force models take into account the tip radius and therefore fitting the data to models seems to us like an appropriate solution. In the same work we compared the amplitude curve methods with the force methods (top and bottom in Figure 15).

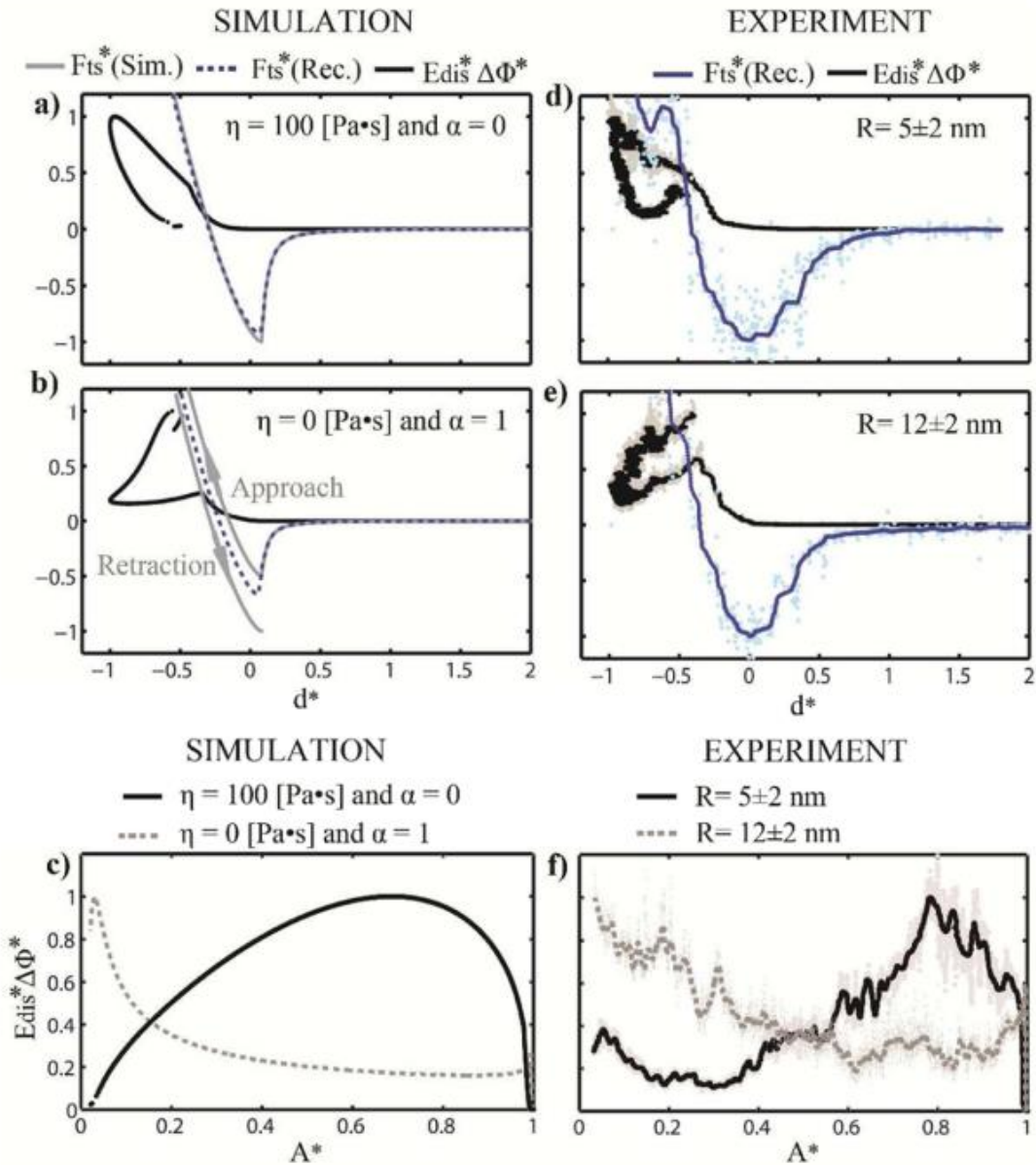

Figure 15. a-c) Simulations and d−f) experiments conducted on a freshly cleaved mica sample where conservative forces and short-range dissipative mechanisms are present. All axes are normalized, and asterisks imply normalization. A transition in the prevalent mechanism of dissipation is observed by simply increasing the tip radius from ≈5 to ≈12 nm, as verified by the $E_{dis}\Delta\phi$ method. The x-axes in a,b) and d,e) are normalized tip−sample distances d* and have been normalized with absolute minima in the corresponding amplitude curves. The x-axes in c) and f) are normalized amplitudes $\bar{A}$ and have been normalized in terms of the corresponding free amplitude $A_0$. Adapted from Ref. 51.

We note that in monomodal dAFM, i.e. where the cantilever is driven at one frequency only, one expression is obtained for conservative interactions[5, 74] and another for dissipative interactions[24, 25]. Conservative forces control the dynamics via what is typically termed the Virial due to the work of San Paulo and Garcia [74] while the dissipative contribution is typically termed dissipation (Eq. 6). The Virial is the time averaged product of the force $F_{ts}$ with the tip position z, i.e. <$F_{ts}z$>, while the energy dissipation $E_{dis}$ is the time average product of $F_{ts}$ with the tip velocity $\dot{z}$, , i.e. < $F_{ts}\dot{z}$>. A combination of these can be exploited to derive [75] the loss tangent tanδ, i.e. loss tan. The expression is independent of the contact area, a point that was raised by Proksch et al. and can be expressed as

$$tan\delta = \frac{<F_{ts}\dot{z}>}{\omega<F_{ts}z>} \qquad (17)$$

where, for the first resonant mode and reducing the motion to the fundamental harmonic, all parameters can be derived from observables independently of $F_{ts}$ by exploiting the equation of motion in (5). Proksch et al. further derived an expression to account for higher harmonics. More recently[76] the following compact (equivalent) expression was proposed and exploited experimentally to discuss lack of reproducibility by invoking the fact that calibration, assumptions in the models, systematic and random error were to be considered,

$$tan\delta = \frac{\Omega\bar{A}-sin\phi}{Q\bar{A}(1-\Omega^2)-cos\phi}$$

(18)

where $\Omega$ is the ratio between the drive $\omega$ and the free resonant frequency $\omega_0$ and $\bar{A}$ is the ratio between the amplitude A and free amplitude $A_0$. Driving at resonance

$$tan\delta = \frac{sin\phi-\bar{A}}{cos\phi}$$

(19)

which is the expression for the first or fundamental harmonic as derived by the authors in 2012 (Eq. 4). As a note regarding the claim of Proksch et al. about tip radius independent, i.e. what they roughly mean by area of interaction, we claim that it is still debatable whether a "sharp" tip probes a sample in the same way that a "blunt" tip does. Therefore, even when a parameter does not depend on the tip, the phenomena in the interaction itself might. In any case, the issue of calibration is recurrent in AFM and does not fully settle even for obvious parameters such as the role of the radius of the tip [20, 35], tip wear[77], spring constant[54] or errors in the calibration of the resonant frequency[76, 78].

## VII. Energy transfer and multifrequency AFM

In order to extract material properties, it is arguably required to get a set of equations that relate the unknowns to expressions. As noted in the previous sections, force/energy models have several unknowns and many times there is a lack of expressions. While monomodal AFM can also be exploited to analyse harmonic distortion[12] one must sometimes hammer the surface. In bimodal AFM two frequencies are excited, one at or near the resonance of the first two, or other, modes. Rodriguez and Garcia first showed in 2004[14] that a second drive near the second mode made quantification[3] and analysis of other frequencies simpler[22, 28] without hammering the sample[79]. As a brief summary, the key point is that rather than access to a single Virial, the Virials of the first $V_1$ and second mode $V_2$ can be quantified in terms of observables, i.e. amplitudes, phases, spring constants, Q factors and resonant frequencies. Energy transfer between the first mode and the second mode is enhanced in bimodal AFM. The consequences are no trivial. For example, even when there is no energy dissipation in the tip-sample junction, i.e. $E_{dis}=0$, the non-linear interaction might lead to a transfer of energy from the first to the second mode and vice-versa by a complex interaction between the harmonics. Also, since there are two drives, it does not necessarily follow that no work can be done on the fundamental frequency, that is $E_{dis}$ at the fundamental frequency might be negative [33].

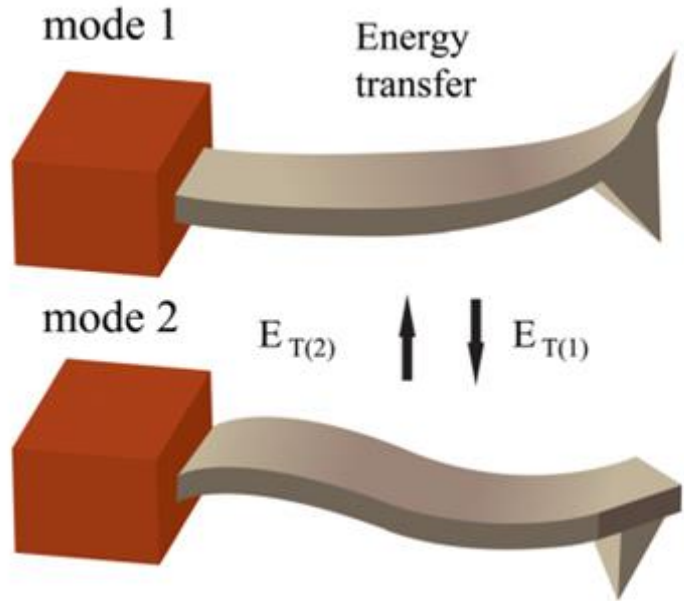

Figure 16. Illustration of the energy transfer between modes in bimodal AFM. Adapted from Ref. 33.

In terms of available expressions, in 2014 Herruzo et al. first showed that with the model in Eq. (16), it was required the sample maximum indentation $\delta_M$ was known to fully determine $\eta$. The extra viral of the second mode $V_2$ would provide an extra equation, that is a total of three when accounting for dissipation through the fundamental frequency Eq. (6), and in that work they exploited a viscoelastic model with three unknowns, i.e. the Young's modulus $E^*$, $\eta$ and $\delta_M$. The explicitly claimed that they ignored long range forces, conservative and dissipative and imaged in the repulsive regime minimizing the attractive contributions [80]. In our experience with long range interaction, i.e. vdW type, it has become clear that one should not assume an inverse square law at least experimentally[81]. Herruzo et al. also developed an expression to test the validity of the assumption, i.e. Hertz contact mechanics, in terms of the two excited frequencies. In 2016 we showed[50] that in the attractive regime the Virial expressions could be exploited to recover the Hamaker H coefficient and assumed an inverse square law. If the conservative expression in Eq. 3 is employed, the unknowns are the minimum distance of approach $d_m$ and H. Arguably the tip radius is also an unknown but we exploited an *insitu* method that we developed to characterize it[42, 82]. We obtained the following expression

$$d_m + bd_m^{2/3} + c = 0 \tag{20}$$

that is, an implicit expression to solve for $d_m$ where b and c are functions of the amplitudes, spring constants, Q factors and phases of the two modes. H is given by

$$H = -\frac{3\pi k_2 A_{02} cos\phi_2}{0.83 R Q_2 A_2} \sqrt{d_m^5 A_1} \tag{21}$$

Where all parameters are known. We did not attempt to use the energy expression to solve for Eqs. (14) and (15), i.e. $\alpha_{nc}$, but the solution is trivial once $d_m$ and H are known by assuming only one dissipative mechanism. To our knowledge such maps have not been shown to date. The expressions by Herruzo et al. have more recently[83] been written more compactly and applied to soft mater. We provide their expressions

$$\delta_M = \left(\frac{A_2^2}{A_1}\right)\left(\frac{V_1}{V_2}\right) \tag{22}$$

$$E^* = \left(4\sqrt{2}k_1 Q_1/\sqrt{RA_1}\right)\left(\frac{k_2}{k_1}\right)^2 \left(\frac{\Delta\omega_2}{\omega_{02}}\right)^2 / cos^2\phi_1 \tag{23}$$

$$\eta_c = (2\pi\omega_1)^{-1} E^* E_{dis}/V_1 \tag{24}$$

where the subscripts 1 and 2 stand for mode 1 and 2. The expressions for the virials in terms of observables are well established[28, 33]. More equations could be obtained, in principle, by exciting other frequencies[84]. Recently, Carlos Álvarez Amo, in the group of Garcia, developed[85] very compact expressions to obtain material properties in the attractive regime for different power laws.

## Conclusions

The evolution of the methods in dAFM shows convergence[86] in the field in that derivations by multiple groups, together with the implementation of the methods[87], are leading to robust quantification and a thorough understanding of the requirements of the technique to become truly analytic[13]. We have shown how material parameters can now be recovered through several methods in dAFM but also how fields of research open in that these show us the complexity of some materials[52, 88]. Improved calibration and the continuous development of analytic solutions should lead to routinely mapping material properties with nanometer and atomic resolution. Continuous models are relevant in that they allow us to parametrize samples through their properties. Still, while the advantage of continuous models should be acknowledged and admitted, it is probably the destiny of the atomic microscope to teach and describe materials by providing a deeper insight of nanometer and atomic processes. After all, friction is not a fundamental force and atomic processes should explain forces and the phenomena that is classically observed [89, 90]. dAFM methods are further advancing in terms of throughput[91], the imaging of cells[92] and sophisticated control and data processing methods[93] implying that the scope of the community is increasing together with an understanding of the nanoscale. Recently Garcia[88] has published a thorough review on the most recent advances on nanomechanical mappings from force distance methods, to force volume, torsional and parametric methods and bimodal AFM. Our purpose is that our review acts as a complementary work that focuses on dAFM and specifically on amplitude modulation AFM and energy dissipation processes. Finally, we have left open the possibility to map and quantify dissipative mechanisms in the long range.